\documentclass[reviewcopy]{elsarticle}

\usepackage[reviewcopy]{adndt}
\usepackage{longtable}

\usepackage{amsmath}

\usepackage{amssymb}

\usepackage{lscape}

\biboptions{square,sort&compress}
\bibpunct[]{[}{]}{,}{n}{}{;}
\begin{document}

\begin{frontmatter}

\journal{Atomic Data and Nuclear Data Tables}


\title{Magnetizabilities of relativistic hydrogenlike atoms \\ in some arbitrary discrete energy eigenstates}

  \author{Patrycja Stefa{\'n}ska\corref{cor1}}
  \ead{pstefanska@mif.pg.gda.pl}

  \cortext[cor1]{Corresponding author.}

  \address{Atomic Physics Division,
Department of Atomic, Molecular and Optical Physics, 
Faculty of Applied Physics and Mathematics, \\
Gda{\'n}sk University of Technology, 
Narutowicza 11/12, 80--233 Gda{\'n}sk, Poland}

\date{\today}

\begin{abstract}  
We present the results of numerical calculations of magnetizability ($\chi$) of the relativistic one-electron atoms with a pointlike, spinless and motionless nuclei of charge $Ze$. Exploiting the analytical formula for $\chi$ recently derived by us [P. Stefa{\'n}ska, 2015], valid for an arbitrary discrete energy eigenstate, we have found the values of the magnetizability for the ground state and for the first and the second set of excited states (i.e.: $2s_{1/2}$, $2p_{1/2}$, $2p_{3/2}$, $3s_{1/2}$, $3p_{1/2}$, $3p_{3/2}$, $3d_{3/2}$, and $3d_{5/2}$) of the Dirac one-electron atom. The results for ions with the atomic number  $1 \leqslant Z \leqslant 137$ are given in 14 tables. The  comparison of the numerical values of magnetizabilities for the ground state and for each states belonging to the first set of excited states of selected hydrogenlike ions, obtained with the use of two different values of the fine-structure constant, i.e.: $\alpha^{-1}=137.035 \: 999 \: 139$ (CODATA 2014) and $\alpha^{-1}=137.035 \: 999 \: 074$ (CODATA 2010), is also presented.\\


\begin{center}
\large{\textbf{Published as: At. Data Nucl. Data Tables 108 (2016) 193--210}
\\*[1ex]
\textbf{doi: 10.1016/j.adt.2015.09.001}\\*[5ex]}
\end{center}
\end{abstract}

\end{frontmatter}




\newpage

\tableofcontents
\listofDtables
\listofDfigures
\vskip5pc


\section{Introduction}

Interaction of atoms and molecules with electromagnetic field is undoubtedly one of the most commonly reported physical processes, both theoretically and experimentally. For the simplest systems, like one-electron atoms, there are some analytical methods of calculating many atomic parameters, such as the polarizability or the magnetizability. One of such useful tool is the Sturmian expansion of the first-order generalized Dirac--Coulomb Green function \cite{Szmy97}, proposed by Szmytkowski in 1997. In the series of papers \cite{Szmy02b,Szmy04,Szmy02a,Miel06,Szmy11,Stef12,Szmy12,Szmy14} published by his group over the period of past several years, it has been used in perturbation-theory calculations of some electromagnetic properties of the relativistic hydrogenlike atoms in the ground state, with a poinlike, spinless and motionless nuclei of charge $Z e$.

Recently, we have shown that the usefulness of this method goes beyond the study of the atomic ground state. In Ref.\ \cite{Stef15} we derived analytically an expression for the magnetizability of the Dirac one-electron atom (with regard to its nucleus we impose the same assumptions as above) in an arbitrary discrete energy eigenstate, characterized by the set of quantum numbers $\{n, \kappa, \mu \}$, in which $n$ denotes the radial quantum number, the Dirac quantum number $\kappa$ is an integer different form zero, whereas $\mu=-|\kappa|+\frac{1}{2}, -|\kappa|+\frac{3}{2}, \ldots, |\kappa|-\frac{1}{2}$ is the magnetic quantum number. The final result has the following form:
\begin{eqnarray}
\chi \equiv \chi_{n\kappa\mu}&=&\frac{\alpha^2 a_0^3}{Z^2}\frac{1}{128(4\kappa^2-1)^2 N_{n \kappa}}
\Bigg\{
\Theta_{n\kappa\mu}^{(\textrm{I})}+\sum_{\kappa'}
\frac{\eta_{\kappa\mu}^{(+)}\delta_{\kappa',-\kappa+1}+\eta_{\kappa\mu}^{(-)} \delta_{\kappa',-\kappa-1}}{N_{n\kappa}+\kappa'}
\nonumber \\
&&\times
\Bigg[
\Theta_{n\kappa}^{(\textrm{II})}+\frac{n!(n^2+2n\gamma_{\kappa}+\kappa^2)\Gamma(n+2\gamma_{\kappa}+1)}{(N_{n\kappa}-\kappa)(\gamma_{\kappa'}-\gamma_{\kappa}-n+1)\Gamma(2\gamma_{\kappa'}+1)}\sum_{k=0}^n\sum_{p=0}^n \widetilde{\mathcal{Z}}_{\kappa \kappa'}^{n}(k)\widetilde{\mathcal{Z}}_{\kappa \kappa'}^{n}(p)
\nonumber \\
&&\quad 
\times
{}_3F_2 
\left(
\begin{array}{c} 
\gamma_{\kappa'}-\gamma_{\kappa}-k-1,\: 
\gamma_{\kappa'}-\gamma_{\kappa}-p-1,\: 
\gamma_{\kappa'}-\gamma_{\kappa}-n+1 \\
\gamma_{\kappa'}-\gamma_{\kappa}-n+2,\:
2\gamma_{\kappa'}+1
\end{array}
;1 
\right) 
\Bigg]
\Bigg\},
\label{eq:1}
\end{eqnarray}
where $\alpha$ is the Sommerfeld's fine structure constant, $\Gamma(\zeta)$ denotes the Euler's gamma function, ${}_3F_2$ is the generalized hypergeometric function, while
\begin{equation}
\eta_{\kappa\mu}^{(\pm)}=(4\kappa^2-1)^2-4\mu^2(2\kappa\pm1)^2,
\label{eq:2}
\end{equation} 
\begin{equation}
\Theta_{n\kappa\mu}^{(\textrm{I})}=-256\kappa^2 \mu^2
\left[
2\kappa^2(n+\gamma_{\kappa})^3+(n+\gamma_{\kappa})(5n^2+10n \gamma_{\kappa}+2\gamma_{\kappa}^2-2\kappa^2+1)N_{n\kappa}^2-\kappa(3n^2+6n \gamma_{\kappa}+4\gamma_{\kappa}^2-\kappa^2)N_{n\kappa}
\right],
\label{eq:3}
\end{equation}
\begin{equation}
\Theta_{n\kappa}^{(\textrm{II})}=2(2n+2\gamma_{\kappa}+1)(\kappa-N_{n \kappa})N_{n\kappa}^2
\left[
5(n+\gamma_{\kappa})(n+\gamma_{\kappa}+1)-3(\gamma_{\kappa}^2-1)
\right],
\label{eq:4}
\end{equation}
\begin{equation}
\widetilde{\mathcal{Z}}_{\kappa \kappa'}^{n}(k)= \frac{(-)^{k} \left[2(N_{n \kappa}-\kappa)+(n-k)(\kappa+\kappa')\right]}{k!(n-k)!} \:
 \frac{\Gamma(\gamma_{\kappa}+\gamma_{\kappa'}+k+2)}{\Gamma(k+2\gamma_{\kappa}+1)}
\label{eq:5}
\end{equation}
and analogously for $\widetilde{\mathcal{Z}}_{\kappa \kappa'}^{n}(p)$, with
\begin{equation}
N_{n\kappa}=\sqrt{n^2+2n\gamma_{\kappa}+\kappa^2},
\label{eq:6}
\end{equation}
and
\begin{equation}
\gamma_{\kappa}=\sqrt{\kappa^2-(\alpha Z)^2}.
\label{eq:7}
\end{equation}

The above result has been exhaustively verified by us, both analytically and numerically. In Ref.\ \cite{Stef15} we have shown that it remains valid for {{\em an arbitrary} discrete energy eigenstate. However, the aforementioned article contains only two representational tables with values of the relativistic magnetizabilities for some excited states of selected hydrogenlike ions. In this work, we present a more comprehensive numerical data in the form of 14 tables comprising the results for the atomic ground state $1s_{1/2}$ and for each state belonging to the first and second set of excited states, i.e.: $2s_{1/2}$, $2p_{1/2}$, $2p_{3/2}$, $3s_{1/2}$, $3p_{1/2}$, $3p_{3/2}$, $3d_{3/2}$ and $3d_{5/2}$, having regard all possible values of the magnetic quantum number. 

Present calculations have been performed with the current value $137.035 \: 999 \: 139$ of the inverse of the fine-structure constant recommended by the Committee on Data for Science and Technology (CODATA) \cite{Mohr14}, in contrast to those described in Ref.\ \cite{Stef15}, in which we have used $\alpha^{-1}=137.035 \: 98 \:95$ (CODATA 1986), to be able to compare our results with the previous results of other authors \cite{Rutk07}. To show how the change in the value of the fine structure constant affects the value of the magnetizability, additionally, we performed calculations for $\chi$ with $\alpha^{-1}=137.035 \: 999 \: 074$, recommended by CODATA 2010 (i.e. immediately before the currently valid value of this constant) and compared them with the corresponding values from Tables \ref{tab:1s_1-2} -- \ref{tab:2p_3-2_mu_3-2}. The appropriate juxtaposition of numerical values of $\chi$ for the ground state and the first excited states of selected hydrogenlike ions are shown in Table \ref{tab:comparison}.



\ack
I wish to thank Professor R.\ Szmytkowski for many stimulating discussions, especially for every advice he gave me during the preparation of this work.



\clearpage

\newpage

\TableExplanation

In all the tables we have used the following notation:

\bigskip
\renewcommand{\arraystretch}{1.0}

\begin{tabular}{ll}
$\chi$ 				& the atomic magnetizability \\
$Z$ 					& the atomic number \\
$\alpha$ 			& the fine structure constant \\
$a_0$					& the Bohr radius \\
$a(b)$				& this means $a \times 10^b$ \\
$\mu$         & the magnetic quantum number; $\mu=-j, \ldots, j$ \\*[1.5ex]
\multicolumn{2}{l}{The atomic states are referred to as $\mathcal{N}{x}_{j}$, due to the following guidelines:} \\*[1ex]
$\mathcal{N}$  & the principal quantum number \\
$\mathcal{N}=n+|\kappa|$ & Note that for the radial quantum number $n=0$, $\kappa$ is the negative integer! \\
$\kappa$ 			& the Dirac quantum number; $\kappa=(l-j)(2j+1)$\\
$j$						& the total angular momentum quantum number; $j=|\kappa|-\frac{1}{2}=l \pm \frac{1}{2}$ \\
$x \equiv s$	& states with the orbital angular momentum quantum number $l=0$ \\
$x \equiv p$	& states with the orbital angular momentum quantum number  $l=1$ \\
$x \equiv d$	& states with the orbital angular momentum quantum number  $l=2$ \\
\end{tabular}

\datatables 



\setlength{\LTleft}{0pt}
\setlength{\LTright}{0pt} 


\setlength{\tabcolsep}{0.5\tabcolsep}

\renewcommand{\arraystretch}{1.0}

\footnotesize 

\begin{longtable}{@{\extracolsep\fill}rrrrrr@{}}
\caption{Relativistic magnetizabilities $\chi$ (in the units of $\alpha^2 a_0^3$) for the ground state  ($1s_{1/2}$) of hydrogenlike atoms, obtained with $\alpha^{-1}=137.035 \: 999 \: 139$ (CODATA 2014).}
\label{tab:1s_1-2}
$Z$ & \multicolumn{1}{c}{$\chi[\alpha^2 a_0^3]$} & $Z$ & \multicolumn{1}{c}{$\chi[\alpha^2 a_0^3]$} & $Z$ & \multicolumn{1}{c}{$\chi[\alpha^2 a_0^3]$} \\
\hline\\
\endfirsthead
\caption[]{(continued)}
$Z$ & \multicolumn{1}{c}{$\chi[\alpha^2 a_0^3]$} & $Z$ & \multicolumn{1}{c}{$\chi[\alpha^2 a_0^3]$} & $Z$ & \multicolumn{1}{c}{$\chi[\alpha^2 a_0^3]$} \\
\hline\\
\endhead
1    &   $-$4.999 644 993 669($-$1)   &   47   &   $-$1.914 539 761 522($-$4)   &   93    &   $-$2.485 921 010 714($-$5)                          \\
2    &   $-$1.249 645 001 764($-$1)   &   48   &   $-$1.821 478 207 432($-$4)   &   94    &   $-$2.369 686 853 502($-$5)                          \\
3    &   $-$5.552 005 708 136($-$2)   &   49   &   $-$1.734 078 279 628($-$4)   &   95    &   $-$2.257 394 374 192($-$5)                          \\
4    &   $-$3.121 450 341 512($-$2)   &   50   &   $-$1.651 892 129 562($-$4)   &   96    &   $-$2.148 886 006 463($-$5)                          \\
5    &   $-$1.996 450 584 458($-$2)   &   51   &   $-$1.574 515 385 043($-$4)   &   97    &   $-$2.044 012 432 389($-$5)                          \\
6    &   $-$1.385 339 770 334($-$2)   &   52   &   $-$1.501 582 182 786($-$4)   &   98    &   $-$1.942 632 090 285($-$5)                          \\
7    &   $-$1.016 859 395 768($-$2)   &   53   &   $-$1.432 760 850 756($-$4)   &   99    &   $-$1.844 610 718 053($-$5)                          \\
8    &   $-$7.777 016 376 682($-$3)   &   54   &   $-$1.367 750 144 935($-$4)   &   100   &   $-$1.749 820 929 318($-$5)                          \\
\hline
\end{longtable}

\newpage

\footnotesize
\begin{longtable}{@{\extracolsep\fill}rrrrrr@{}}
\caption{Relativistic magnetizabilities $\chi$ (in the units of $\alpha^2 a_0^3$)  of hydrogenlike atoms in the excited state $2s_{1/2}$, obtained with $\alpha^{-1}=137.035 \: 999 \: 139$ (CODATA 2014).}
\label{tab:2s_1-2}
$Z$ & \multicolumn{1}{c}{$\chi[\alpha^2 a_0^3]$} & $Z$ & \multicolumn{1}{c}{$\chi[\alpha^2 a_0^3]$} & $Z$ & \multicolumn{1}{c}{$\chi[\alpha^2 a_0^3]$} \\
\hline\\
\endfirsthead
\caption[]{(continued)}
$Z$ & \multicolumn{1}{c}{$\chi[\alpha^2 a_0^3]$} & $Z$ & \multicolumn{1}{c}{$\chi[\alpha^2 a_0^3]$} & $Z$ & \multicolumn{1}{c}{$\chi[\alpha^2 a_0^3]$} \\
\hline\\
\endhead
1    &   $-$6.999 722 649 147($+$0)    &   47   &   $-$2.891 300 079 244($-$3)   &   93    &   $-$5.295 132 413 795($-$4)                             \\
2    &   $-$1.749 722 649 002($+$0)    &   48   &   $-$2.760 626 290 162($-$3)   &   94    &   $-$5.122 608 926 758($-$4)                             \\
3    &   $-$7.775 004 265 370($-$1)   &   49   &   $-$2.637 869 458 376($-$3)   &   95    &   $-$4.955 404 845 206($-$4)                             \\
4    &   $-$4.372 226 484 179($-$1)   &   50   &   $-$2.522 402 512 753($-$3)   &   96    &   $-$4.793 293 766 449($-$4)                             \\
5    &   $-$2.797 226 479 764($-$1)   &   51   &   $-$2.413 659 246 069($-$3)   &   97    &   $-$4.636 060 481 284($-$4)                             \\
6    &   $-$1.941 670 918 773($-$1)   &   52   &   $-$2.311 127 360 473($-$3)   &   98    &   $-$4.483 500 254 287($-$4)                             \\
7    &   $-$1.425 797 896 419($-$1)   &   53   &   $-$2.214 342 422 647($-$3)   &   99    &   $-$4.335 418 150 281($-$4)                             \\
8    &   $-$1.090 976 460 292($-$1)   &   54   &   $-$2.122 882 595 177($-$3)   &   100   &   $-$4.191 628 402 646($-$4)                             \\
\hline
\end{longtable}

\newpage

\footnotesize
\begin{longtable}{@{\extracolsep\fill}rrrrrr@{}}
\caption{Relativistic magnetizabilities $\chi$ (in the units of $\alpha^2 a_0^3$)  of hydrogenlike atoms in the excited state $2p_{1/2}$, obtained with $\alpha^{-1}=137.035 \: 999 \: 139$ (CODATA 2014).}
\label{tab:2p_1-2}
$Z$ & \multicolumn{1}{c}{$\chi[\alpha^2 a_0^3]$} & $Z$ & \multicolumn{1}{c}{$\chi[\alpha^2 a_0^3]$} & $Z$ & \multicolumn{1}{c}{$\chi[\alpha^2 a_0^3]$} \\
\hline\\
\endfirsthead
\caption[]{(continued)}
$Z$ & \multicolumn{1}{c}{$\chi[\alpha^2 a_0^3]$} & $Z$ & \multicolumn{1}{c}{$\chi[\alpha^2 a_0^3]$} & $Z$ & \multicolumn{1}{c}{$\chi[\alpha^2 a_0^3]$} \\
\hline\\
\endhead
1    &   6.676 163 152 752($+$4)    &   47   &   1.041 492 649 963($-$2)   &   93    &   2.075 091 052 946($-$4)                              \\
2    &   4.171 164 660 772($+$3)    &   48   &   9.452 888 853 908($-$3)   &   94    &   1.888 736 138 669($-$4)                              \\
3    &   8.234 606 720 151($+$2)    &   49   &   8.591 529 248 763($-$3)   &   95    &   1.715 865 142 938($-$4)                              \\
4    &   2.603 386 065 977($+$2)    &   50   &   7.818 803 550 735($-$3)   &   96    &   1.555 531 822 422($-$4)                              \\
5    &   1.065 244 131 032($+$2)    &   51   &   7.124 297 098 369($-$3)   &   97    &   1.406 862 128 024($-$4)                              \\
6    &   5.130 676 007 305($+$1)    &   52   &   6.498 978 907 081($-$3)   &   98    &   1.269 048 169 139($-$4)                              \\
7    &   2.765 269 415 830($+$1)    &   53   &   5.934 996 687 137($-$3)   &   99    &   1.141 342 734 163($-$4)                              \\
8    &   1.618 150 944 960($+$1)    &   54   &   5.425 505 295 092($-$3)   &   100   &   1.023 054 311 643($-$4)                              \\
\hline
\end{longtable}

\newpage

\footnotesize
\begin{longtable}{@{\extracolsep\fill}rrrrrr@{}}
\caption{Relativistic magnetizabilities $\chi$ (in the units of $\alpha^2 a_0^3$)  of hydrogenlike atoms in the excited state $2p_{3/2}$ ($\mu=\pm 1/2$), obtained with $\alpha^{-1}=137.035 \: 999 \: 139$ (CODATA 2014).}
\label{tab:2p_3-2_mu_1-2}
$Z$ & \multicolumn{1}{c}{$\chi[\alpha^2 a_0^3]$} & $Z$ & \multicolumn{1}{c}{$\chi[\alpha^2 a_0^3]$} & $Z$ & \multicolumn{1}{c}{$\chi[\alpha^2 a_0^3]$} \\
\hline\\
\endfirsthead
\caption[]{(continued)}
$Z$ & \multicolumn{1}{c}{$\chi[\alpha^2 a_0^3]$} & $Z$ & \multicolumn{1}{c}{$\chi[\alpha^2 a_0^3]$} & $Z$ & \multicolumn{1}{c}{$\chi[\alpha^2 a_0^3]$} \\
\hline\\
\endhead
1    &   $-$6.677 063 123 774($+$4)    &   47   &   $-$1.419 858 008 271($-$2)   &   93    &   $-$9.520 247 585 051($-$4)                             \\
2    &   $-$4.173 414 370 995($+$3)    &   48   &   $-$1.306 850 716 417($-$2)   &   94    &   $-$9.110 967 976 025($-$4)                             \\
3    &   $-$8.244 603 822 364($+$2)    &   49   &   $-$1.204 929 044 477($-$2)   &   95    &   $-$8.721 966 496 674($-$4)                             \\
4    &   $-$2.609 008 168 169($+$2)    &   50   &   $-$1.112 807 953 802($-$2)   &   96    &   $-$8.352 003 098 775($-$4)                             \\
5    &   $-$1.068 841 233 198($+$2)    &   51   &   $-$1.029 373 178 218($-$2)   &   97    &   $-$7.999 924 087 903($-$4)                             \\
6    &   $-$5.155 647 028 648($+$1)    &   52   &   $-$9.536 557 503 502($-$3)   &   98    &   $-$7.664 655 136 787($-$4)                             \\
7    &   $-$2.783 607 783 733($+$1)    &   53   &   $-$8.848 107 262 219($-$3)   &   99    &   $-$7.345 194 910 790($-$4)                             \\
8    &   $-$1.632 184 465 485($+$1)    &   54   &   $-$8.220 993 530 005($-$3)   &   100   &   $-$7.040 609 243 754($-$4)                             \\
\hline
\end{longtable}

\newpage

\footnotesize
\begin{longtable}{@{\extracolsep\fill}rrrrrr@{}}
\caption{Relativistic magnetizabilities $\chi$ (in the units of $\alpha^2 a_0^3$)  of hydrogenlike atoms in the excited state $2p_{3/2}$ ($\mu=\pm 3/2$), obtained with $\alpha^{-1}=137.035 \: 999 \: 139$ (CODATA 2014).}
\label{tab:2p_3-2_mu_3-2}
$Z$ & \multicolumn{1}{c}{$\chi[\alpha^2 a_0^3]$} & $Z$ & \multicolumn{1}{c}{$\chi[\alpha^2 a_0^3]$} & $Z$ & \multicolumn{1}{c}{$\chi[\alpha^2 a_0^3]$} \\
\hline\\
\endfirsthead
\caption[]{(continued)}
$Z$ & \multicolumn{1}{c}{$\chi[\alpha^2 a_0^3]$} & $Z$ & \multicolumn{1}{c}{$\chi[\alpha^2 a_0^3]$} & $Z$ & \multicolumn{1}{c}{$\chi[\alpha^2 a_0^3]$} \\
\hline\\
\endhead
1    &   $-$5.999 861 546 894($+$0)    &   47   &   $-$2.578 631 244 428($-$3)   &   93    &   $-$5.589 508 155 663($-$4)   \\
2    &   $-$1.499 861 548 141($+$0)    &   48   &   $-$2.466 676 723 115($-$3)   &   94    &   $-$5.443 509 574 534($-$4)   \\
3    &   $-$6.665 282 168 857($-$1)   &   49   &   $-$2.361 509 657 648($-$3)   &   95    &   $-$5.302 132 488 766($-$4)   \\
4    &   $-$3.748 615 531 285($-$1)   &   50   &   $-$2.262 592 588 983($-$3)   &   96    &   $-$5.165 185 776 675($-$4)   \\
5    &   $-$2.398 615 568 695($-$1)   &   51   &   $-$2.169 440 228 389($-$3)   &   97    &   $-$5.032 488 119 440($-$4)   \\
6    &   $-$1.665 282 281 087($-$1)   &   52   &   $-$2.081 613 496 469($-$3)   &   98    &   $-$4.903 867 404 056($-$4)   \\
7    &   $-$1.223 105 464 382($-$1)   &   53   &   $-$1.998 714 341 935($-$3)   &   99    &   $-$4.779 160 168 287($-$4)   \\
8    &   $-$9.361 157 308 251($-$2)   &   54   &   $-$1.920 381 225 688($-$3)   &   100   &   $-$4.658 211 084 276($-$4)   \\
\hline
\end{longtable}

\newpage

\footnotesize
\begin{longtable}{@{\extracolsep\fill}rrrrrr@{}}
\caption{Relativistic magnetizabilities $\chi$ (in the units of $\alpha^2 a_0^3$)  of hydrogenlike atoms in the excited state $3s_{1/2}$, obtained with $\alpha^{-1}=137.035 \: 999 \: 139$ (CODATA 2014).}
\label{tab:3s_1-2}
$Z$ & \multicolumn{1}{c}{$\chi[\alpha^2 a_0^3]$} & $Z$ & \multicolumn{1}{c}{$\chi[\alpha^2 a_0^3]$} & $Z$ & \multicolumn{1}{c}{$\chi[\alpha^2 a_0^3]$} \\
\hline\\
\endfirsthead
\caption[]{(continued)}
$Z$ & \multicolumn{1}{c}{$\chi[\alpha^2 a_0^3]$} & $Z$ & \multicolumn{1}{c}{$\chi[\alpha^2 a_0^3]$} & $Z$ & \multicolumn{1}{c}{$\chi[\alpha^2 a_0^3]$} \\
\hline\\
\endhead
1    &   $-$3.449 899 704 639($+$1)    &   47   &   $-$1.460 589 595 746($-$2)   &   93    &   $-$2.938 787 161 517($-$3)                            \\
2    &   $-$8.623 997 035 001($+$0)    &   48   &   $-$1.396 150 304 355($-$2)   &   94    &   $-$2.852 875 528 084($-$3)                            \\
3    &   $-$3.832 330 349 350($+$0)    &   49   &   $-$1.335 612 093 153($-$2)   &   95    &   $-$2.769 568 453 095($-$3)                            \\
4    &   $-$2.155 246 989 428($+$0)    &   50   &   $-$1.278 665 898 809($-$2)   &   96    &   $-$2.688 753 878 288($-$3)                            \\
5    &   $-$1.378 996 955 223($+$0)    &   51   &   $-$1.225 032 654 958($-$2)   &   97    &   $-$2.610 325 233 432($-$3)                            \\
6    &   $-$9.573 302 467 217($-$1)   &   52   &   $-$1.174 459 864 603($-$2)   &   98    &   $-$2.534 181 079 133($-$3)                            \\
7    &   $-$7.030 784 965 588($-$1)   &   53   &   $-$1.126 718 620 861($-$2)   &   99    &   $-$2.460 224 772 109($-$3)                            \\
8    &   $-$5.380 593 067 547($-$1)   &   54   &   $-$1.081 601 010 268($-$2)   &   100   &   $-$2.388 364 150 764($-$3)                            \\
\hline
\end{longtable}

\newpage

\footnotesize
\begin{longtable}{@{\extracolsep\fill}rrrrrr@{}}
\caption{Relativistic magnetizabilities $\chi$ (in the units of $\alpha^2 a_0^3$)  of hydrogenlike atoms in the excited state $3p_{1/2}$, obtained with $\alpha^{-1}=137.035 \: 999 \: 139$ (CODATA 2014).}
\label{tab:3p_1-2}
$Z$ & \multicolumn{1}{c}{$\chi[\alpha^2 a_0^3]$} & $Z$ & \multicolumn{1}{c}{$\chi[\alpha^2 a_0^3]$} & $Z$ & \multicolumn{1}{c}{$\chi[\alpha^2 a_0^3]$} \\
\hline\\
\endfirsthead
\caption[]{(continued)}
$Z$ & \multicolumn{1}{c}{$\chi[\alpha^2 a_0^3]$} & $Z$ & \multicolumn{1}{c}{$\chi[\alpha^2 a_0^3]$} & $Z$ & \multicolumn{1}{c}{$\chi[\alpha^2 a_0^3]$} \\
\hline\\
\endhead
1    &   2.253 079 648 737($+$5)    &   47   &   2.961 501 640 474($-$2)    &   93    &   $-$5.821 665 076 994($-$4)                           \\
2    &   1.407 454 544 846($+$4)    &   48   &   2.660 256 095 643($-$2)    &   94    &   $-$6.134 954 630 563($-$4)                           \\
3    &   2.777 786 384 569($+$3)    &   49   &   2.391 571 609 084($-$2)    &   95    &   $-$6.412 175 427 710($-$4)                           \\
4    &   8.778 590 781 585($+$2)    &   50   &   2.151 495 639 188($-$2)    &   96    &   $-$6.656 114 912 087($-$4)                           \\
5    &   3.590 183 626 425($+$2)    &   51   &   1.936 614 686 446($-$2)    &   97    &   $-$6.869 336 974 364($-$4)                           \\
6    &   1.728 120 707 372($+$2)    &   52   &   1.743 972 583 622($-$2)    &   98    &   $-$7.054 201 043 586($-$4)                           \\
7    &   9.307 202 556 408($+$1)    &   53   &   1.571 002 397 351($-$2)    &   99    &   $-$7.212 879 385 637($-$4)                           \\
8    &   5.441 676 605 826($+$1)    &   54   &   1.415 469 474 468($-$2)    &   100   &   $-$7.347 372 789 149($-$4)                           \\
\hline
\end{longtable}

\newpage

\footnotesize
\begin{longtable}{@{\extracolsep\fill}rrrrrr@{}}
\caption{Relativistic magnetizabilities $\chi$ (in the units of $\alpha^2 a_0^3$)  of hydrogenlike atoms in the excited state $3p_{3/2}$ ($\mu=\pm 1/2$), obtained with $\alpha^{-1}=137.035 \: 999 \: 139$ (CODATA 2014).}
\label{tab:3p_3-2_mu_1-2}
$Z$ & \multicolumn{1}{c}{$\chi[\alpha^2 a_0^3]$} & $Z$ & \multicolumn{1}{c}{$\chi[\alpha^2 a_0^3]$} & $Z$ & \multicolumn{1}{c}{$\chi[\alpha^2 a_0^3]$} \\
\hline\\
\endfirsthead
\caption[]{(continued)}
$Z$ & \multicolumn{1}{c}{$\chi[\alpha^2 a_0^3]$} & $Z$ & \multicolumn{1}{c}{$\chi[\alpha^2 a_0^3]$} & $Z$ & \multicolumn{1}{c}{$\chi[\alpha^2 a_0^3]$} \\
\hline\\
\endhead
1    &   $-$2.253 619 636 543($+$5)    &   47   &   $-$5.283 002 601 438($-$2)   &   93    &   $-$4.382 984 551 435($-$3)                             \\
2    &   $-$1.408 804 422 899($+$4)    &   48   &   $-$4.880 910 212 206($-$2)   &   94    &   $-$4.217 586 852 505($-$3)                             \\
3    &   $-$2.783 785 165 072($+$3)    &   49   &   $-$4.517 485 297 588($-$2)   &   95    &   $-$4.059 881 012 262($-$3)                             \\
4    &   $-$8.812 328 586 296($+$2)    &   50   &   $-$4.188 291 569 502($-$2)   &   96    &   $-$3.909 413 039 128($-$3)                             \\
5    &   $-$3.611 771 430 726($+$2)    &   51   &   $-$3.889 478 733 899($-$2)   &   97    &   $-$3.765 759 884 777($-$3)                             \\
6    &   $-$1.743 108 511 171($+$2)    &   52   &   $-$3.617 695 415 994($-$2)   &   98    &   $-$3.628 526 974 695($-$3)                             \\
7    &   $-$9.417 284 670 092($+$1)    &   53   &   $-$3.370 016 398 695($-$2)   &   99    &   $-$3.497 345 952 933($-$3)                             \\
8    &   $-$5.525 929 631 019($+$1)    &   54   &   $-$3.143 881 604 543($-$2)   &   100   &   $-$3.371 872 619 588($-$3)                             \\
\hline
\end{longtable}

\newpage

\footnotesize
\begin{longtable}{@{\extracolsep\fill}rrrrrr@{}}
\caption{Relativistic magnetizabilities $\chi$ (in the units of $\alpha^2 a_0^3$)  of hydrogenlike atoms in the excited state $3p_{3/2}$ ($\mu=\pm 3/2$), obtained with $\alpha^{-1}=137.035 \: 999 \: 139$ (CODATA 2014).}
\label{tab:3p_3-2_mu_3-2}
$Z$ & \multicolumn{1}{c}{$\chi[\alpha^2 a_0^3]$} & $Z$ & \multicolumn{1}{c}{$\chi[\alpha^2 a_0^3]$} & $Z$ & \multicolumn{1}{c}{$\chi[\alpha^2 a_0^3]$} \\
\hline\\
\endfirsthead
\caption[]{(continued)}
$Z$ & \multicolumn{1}{c}{$\chi[\alpha^2 a_0^3]$} & $Z$ & \multicolumn{1}{c}{$\chi[\alpha^2 a_0^3]$} & $Z$ & \multicolumn{1}{c}{$\chi[\alpha^2 a_0^3]$} \\
\hline\\
\endhead
1    &   $-$3.599 951 541 316($+$1)    &   47   &   $-$1.581 344 317 096($-$2)   &   93    &   $-$3.681 907 779 187($-$3)   \\
2    &   $-$8.999 515 414 600($+$0)    &   48   &   $-$1.514 152 196 291($-$2)   &   94    &   $-$3.593 908 592 141($-$3)   \\
3    &   $-$3.999 515 417 007($+$0)    &   49   &   $-$1.451 032 127 476($-$2)   &   95    &   $-$3.508 677 457 715($-$3)   \\
4    &   $-$2.249 515 420 376($+$0)    &   50   &   $-$1.391 661 634 388($-$2)   &   96    &   $-$3.426 099 682 411($-$3)   \\
5    &   $-$1.439 515 424 709($+$0)    &   51   &   $-$1.335 749 542 934($-$2)   &   97    &   $-$3.346 066 454 083($-$3)   \\
6    &   $-$9.995 154 300 037($-$1)   &   52   &   $-$1.283 032 404 604($-$2)   &   98    &   $-$3.268 474 483 707($-$3)   \\
7    &   $-$7.342 093 138 125($-$1)   &   53   &   $-$1.233 271 387 727($-$2)   &   99    &   $-$3.193 225 672 345($-$3)   \\
8    &   $-$5.620 154 434 822($-$1)   &   54   &   $-$1.186 249 567 930($-$2)   &   100   &   $-$3.120 226 801 306($-$3)   \\
\hline
\end{longtable}

\newpage

\footnotesize
\begin{longtable}{@{\extracolsep\fill}rrrrrr@{}}
\caption{Relativistic magnetizabilities $\chi$ (in the units of $\alpha^2 a_0^3$)  of hydrogenlike atoms in the excited state $3d_{3/2}$ ($\mu=\pm 1/2$), obtained with $\alpha^{-1}=137.035 \: 999 \: 139$ (CODATA 2014).}
\label{tab:3d_3-2_mu_1-2}
$Z$ & \multicolumn{1}{c}{$\chi[\alpha^2 a_0^3]$} & $Z$ & \multicolumn{1}{c}{$\chi[\alpha^2 a_0^3]$} & $Z$ & \multicolumn{1}{c}{$\chi[\alpha^2 a_0^3]$} \\
\hline\\
\endfirsthead
\caption[]{(continued)}
$Z$ & \multicolumn{1}{c}{$\chi[\alpha^2 a_0^3]$} & $Z$ & \multicolumn{1}{c}{$\chi[\alpha^2 a_0^3]$} & $Z$ & \multicolumn{1}{c}{$\chi[\alpha^2 a_0^3]$} \\
\hline\\
\endhead
1    &   7.300 965 637 634($+$5)    &   47   &   1.382 152 566 398($-$1)   &   93    &   7.014 861 885 531($-$3)                        \\
2    &   4.562 621 482 550($+$4)    &   48   &   1.266 106 404 598($-$1)   &   94    &   6.668 966 264 598($-$3)                        \\
3    &   9.010 998 791 784($+$3)    &   49   &   1.161 726 868 210($-$1)   &   95    &   6.342 269 233 241($-$3)                        \\
4    &   2.850 433 485 628($+$3)    &   50   &   1.067 644 863 689($-$1)   &   96    &   6.033 549 850 614($-$3)                        \\
5    &   1.167 167 467 288($+$3)    &   51   &   9.826 745 342 003($-$2)   &   97    &   5.741 675 544 197($-$3)                        \\
6    &   5.626 520 147 653($+$2)    &   52   &   9.057 858 044 552($-$2)   &   98    &   5.465 594 924 698($-$3)                        \\
7    &   3.035 664 451 768($+$2)    &   53   &   8.360 814 639 921($-$2)   &   99    &   5.204 331 249 137($-$3)                        \\
8    &   1.778 510 190 602($+$2)    &   54   &   7.727 779 708 599($-$2)   &   100   &   4.956 976 467 939($-$3)                        \\
\hline
\end{longtable}

\newpage

\footnotesize
\begin{longtable}{@{\extracolsep\fill}rrrrrr@{}}
\caption{Relativistic magnetizabilities $\chi$ (in the units of $\alpha^2 a_0^3$)  of hydrogenlike atoms in the excited state $3d_{3/2}$ ($\mu=\pm 3/2$), obtained with $\alpha^{-1}=137.035 \: 999 \: 139$ (CODATA 2014).}
\label{tab:3d_3-2_mu_3-2}
$Z$ & \multicolumn{1}{c}{$\chi[\alpha^2 a_0^3]$} & $Z$ & \multicolumn{1}{c}{$\chi[\alpha^2 a_0^3]$} & $Z$ & \multicolumn{1}{c}{$\chi[\alpha^2 a_0^3]$} \\
\hline\\
\endfirsthead
\caption[]{(continued)}
$Z$ & \multicolumn{1}{c}{$\chi[\alpha^2 a_0^3]$} & $Z$ & \multicolumn{1}{c}{$\chi[\alpha^2 a_0^3]$} & $Z$ & \multicolumn{1}{c}{$\chi[\alpha^2 a_0^3]$} \\
\hline\\
\endhead
1    &   4.867 170 427 249($+$5)    &   47   &   8.602 099 797 280($-$2)   &   93    &   3.270 743 993 589($-$3)   \\
2    &   3.041 397 676 633($+$4)    &   48   &   7.854 587 339 729($-$2)   &   94    &   3.074 334 771 839($-$3)   \\
3    &   6.005 777 188 294($+$3)    &   49   &   7.183 268 773 333($-$2)   &   95    &   2.889 647 729 010($-$3)   \\
4    &   1.899 414 206 410($+$3)    &   50   &   6.579 142 215 135($-$2)   &   96    &   2.715 913 505 579($-$3)   \\
5    &   7.775 518 608 467($+$2)    &   51   &   6.034 415 187 203($-$2)   &   97    &   2.552 419 366 998($-$3)   \\
6    &   3.747 126 702 718($+$2)    &   52   &   5.542 322 973 071($-$2)   &   98    &   2.398 504 552 941($-$3)   \\
7    &   2.020 921 318 113($+$2)    &   53   &   5.096 977 048 562($-$2)   &   99    &   2.253 556 048 873($-$3)   \\
8    &   1.183 488 120 139($+$2)    &   54   &   4.693 238 160 067($-$2)   &   100   &   2.117 004 737 932($-$3)   \\
\hline
\end{longtable}

\newpage

\footnotesize
\begin{longtable}{@{\extracolsep\fill}rrrrrr@{}}
\caption{Relativistic magnetizabilities $\chi$ (in the units of $\alpha^2 a_0^3$)  of hydrogenlike atoms in the excited state $3d_{5/2}$ ($\mu=\pm 1/2$), obtained with $\alpha^{-1}=137.035 \: 999 \: 139$ (CODATA 2014).}
\label{tab:3d_5-2_mu_1-2}
$Z$ & \multicolumn{1}{c}{$\chi[\alpha^2 a_0^3]$} & $Z$ & \multicolumn{1}{c}{$\chi[\alpha^2 a_0^3]$} & $Z$ & \multicolumn{1}{c}{$\chi[\alpha^2 a_0^3]$} \\
\hline\\
\endfirsthead
\caption[]{(continued)}
$Z$ & \multicolumn{1}{c}{$\chi[\alpha^2 a_0^3]$} & $Z$ & \multicolumn{1}{c}{$\chi[\alpha^2 a_0^3]$} & $Z$ & \multicolumn{1}{c}{$\chi[\alpha^2 a_0^3]$} \\
\hline\\
\endhead
1    &   $-$7.301 295 633 993($+$5)    &   47   &   $-$1.527 906 889 615($-$1)   &   93    &   $-$1.046 863 644 964($-$2)   \\
2    &   $-$4.563 446 446 141($+$4)    &   48   &   $-$1.405 701 300 375($-$1)   &   94    &   $-$1.004 204 115 620($-$2)   \\
3    &   $-$9.014 665 094 365($+$3)    &   49   &   $-$1.295 535 604 542($-$1)   &   95    &   $-$9.637 181 328 664($-$3)   \\
4    &   $-$2.852 495 621 544($+$3)    &   50   &   $-$1.196 011 148 224($-$1)   &   96    &   $-$9.252 730 884 592($-$3)   \\
5    &   $-$1.168 487 103 207($+$3)    &   51   &   $-$1.105 915 383 609($-$1)   &   97    &   $-$8.887 457 500 952($-$3)   \\
6    &   $-$5.635 683 173 539($+$2)    &   52   &   $-$1.024 194 085 943($-$1)   &   98    &   $-$8.540 215 100 647($-$3)   \\
7    &   $-$3.042 395 504 902($+$2)    &   53   &   $-$9.499 281 518 380($-$2)   &   99    &   $-$8.209 937 010 306($-$3)   \\
8    &   $-$1.783 662 799 902($+$2)    &   54   &   $-$8.823 141 545 890($-$2)   &   100   &   $-$7.895 629 723 348($-$3)   \\
\hline
\end{longtable}

\newpage

\footnotesize
\begin{longtable}{@{\extracolsep\fill}rrrrrr@{}}
\caption{Relativistic magnetizabilities $\chi$ (in the units of $\alpha^2 a_0^3$)  of hydrogenlike atoms in the excited state $3d_{5/2}$ ($\mu=\pm 3/2$), obtained with $\alpha^{-1}=137.035 \: 999 \: 139$ (CODATA 2014).}
\label{tab:3d_5-2_mu_3-2}
$Z$ & \multicolumn{1}{c}{$\chi[\alpha^2 a_0^3]$} & $Z$ & \multicolumn{1}{c}{$\chi[\alpha^2 a_0^3]$} & $Z$ & \multicolumn{1}{c}{$\chi[\alpha^2 a_0^3]$} \\
\hline\\
\endfirsthead
\caption[]{(continued)}
$Z$ & \multicolumn{1}{c}{$\chi[\alpha^2 a_0^3]$} & $Z$ & \multicolumn{1}{c}{$\chi[\alpha^2 a_0^3]$} & $Z$ & \multicolumn{1}{c}{$\chi[\alpha^2 a_0^3]$} \\
\hline\\
\endhead
1    &   $-$4.867 620 421 635($+$5)    &   47   &   $-$1.058 323 534 299($-$1)   &   93    &   $-$7.918 361 781 247($-$3)   \\
2    &   $-$3.042 522 620 491($+$4)    &   48   &   $-$9.751 733 756 547($-$2)   &   94    &   $-$7.611 972 956 068($-$3)   \\
3    &   $-$6.010 776 626 879($+$3)    &   49   &   $-$9.001 516 147 331($-$2)   &   95    &   $-$7.320 766 554 808($-$3)   \\
4    &   $-$1.902 226 145 000($+$3)    &   50   &   $-$8.323 177 537 031($-$2)   &   96    &   $-$7.043 829 849 743($-$3)   \\
5    &   $-$7.793 512 994 430($+$2)    &   51   &   $-$7.708 561 480 069($-$2)   &   97    &   $-$6.780 314 089 776($-$3)   \\
6    &   $-$3.759 621 088 759($+$2)    &   52   &   $-$7.150 573 949 303($-$2)   &   98    &   $-$6.529 429 401 888($-$3)   \\
7    &   $-$2.030 099 377 716($+$2)    &   53   &   $-$6.643 027 880 940($-$2)   &   99    &   $-$6.290 440 146 415($-$3)   \\
8    &   $-$1.190 513 756 379($+$2)    &   54   &   $-$6.180 512 865 865($-$2)   &   100   &   $-$6.062 660 681 625($-$3)   \\
\hline
\end{longtable}

\newpage

\footnotesize
\begin{longtable}{@{\extracolsep\fill}rrrrrr@{}}
\caption{Relativistic magnetizabilities $\chi$ (in the units of $\alpha^2 a_0^3$)  of hydrogenlike atoms in the excited state $3d_{5/2}$ ($\mu=\pm 5/2$), obtained with $\alpha^{-1}=137.035 \: 999 \: 139$ (CODATA 2014).}
\label{tab:3d_5-2_mu_5-2}
$Z$ & \multicolumn{1}{c}{$\chi[\alpha^2 a_0^3]$} & $Z$ & \multicolumn{1}{c}{$\chi[\alpha^2 a_0^3]$} & $Z$ & \multicolumn{1}{c}{$\chi[\alpha^2 a_0^3]$} \\
\hline\\
\endfirsthead
\caption[]{(continued)}
$Z$ & \multicolumn{1}{c}{$\chi[\alpha^2 a_0^3]$} & $Z$ & \multicolumn{1}{c}{$\chi[\alpha^2 a_0^3]$} & $Z$ & \multicolumn{1}{c}{$\chi[\alpha^2 a_0^3]$} \\
\hline\\
\endhead
1    &   $-$2.699 969 190 335($+$1)    &   47   &   $-$1.191 568 236 683($-$2)   &   93    &   $-$2.817 812 444 465($-$3)   \\
2    &   $-$6.749 691 904 782($+$0)    &   48   &   $-$1.141 175 262 133($-$2)   &   94    &   $-$2.751 836 555 806($-$3)   \\
3    &   $-$2.999 691 907 161($+$0)    &   49   &   $-$1.093 836 351 152($-$2)   &   95    &   $-$2.687 937 007 096($-$3)   \\
4    &   $-$1.687 191 910 493($+$0)    &   50   &   $-$1.049 309 646 613($-$2)   &   96    &   $-$2.626 027 780 046($-$3)   \\
5    &   $-$1.079 691 914 777($+$0)    &   51   &   $-$1.007 376 768 019($-$2)   &   97    &   $-$2.566 027 267 424($-$3)   \\
6    &   $-$7.496 919 200 131($-$1)   &   52   &   $-$9.678 401 290 602($-$3)   &   98    &   $-$2.507 858 004 372($-$3)   \\
7    &   $-$5.507 123 343 646($-$1)   &   53   &   $-$9.305 206 060 602($-$3)   &   99    &   $-$2.451 446 418 634($-$3)   \\
8    &   $-$4.215 669 333 417($-$1)   &   54   &   $-$8.952 555 058 160($-$3)   &   100   &   $-$2.396 722 598 180($-$3)   \\
\hline
\end{longtable}

\newpage

\footnotesize

\begin{landscape}
\begin{longtable}{@{\extracolsep\fill}rrrrrr@{}}
\caption{Relativistic magnetizabilities $\chi$ (in the units of $\alpha^2 a_0^3$) for states $1s_{1/2}$, $2s_{1/2}$, $2p_{1/2}$ and $2p_{3/2}$ of selected  hydrogenlike ions, obtained with $\alpha^{-1}=137.035 \: 999 \: 139$ (CODATA 2014) [the upper entries] and with $\alpha^{-1}=137.035 \: 999 \: 074$ (CODATA 2010) [the lower entries].}
\label{tab:comparison}
$Z$ & \multicolumn{1}{c}{$1s_{1/2}$} &  \multicolumn{1}{c}{$2s_{1/2}$} &  \multicolumn{1}{c}{$2p_{1/2}$} & \multicolumn{1}{c}{$2p_{3/2}$ ($\mu=\pm 1/2$)} & \multicolumn{1}{c}{$2p_{3/2}$ ($\mu=\pm 3/2$)}  \\
\hline\\
\endfirsthead
\caption[]{(continued)}
$Z$ & \multicolumn{1}{c}{$\chi[\alpha^2 a_0^3]$} & $Z$ & \multicolumn{1}{c}{$\chi[\alpha^2 a_0^3]$} & $Z$ & \multicolumn{1}{c}{$\chi[\alpha^2 a_0^3]$} \\
\hline\\
\endhead
1    & $-$4.999 644 993 669($-$1) & $-$6.999 722 649 147($+$0) & 6.676 163 152 752($+$4) & $-$6.677 063 123 774($+$4) & $-$5.999 861 546 894($+$0) \\
		 & $-$4.999 644 993 668($-$1) & $-$6.999 722 649 147($+$0) & 6.676 163 146 418($+$4) & $-$6.677 063 117 440($+$4) & $-$5.999 861 546 894($+$0) \\*[2ex] 
2    & $-$1.249 645 001 764($-$1) & $-$1.749 722 649 002($+$0) & 4.171 164 660 772($+$3) & $-$4.173 414 370 995($+$3) & $-$1.499 861 548 141($+$0) \\
     & $-$1.249 645 001 764($-$1) & $-$1.749 722 649 002($+$0) & 4.171 164 656 814($+$3) & $-$4.173 414 367 036($+$3) & $-$1.499 861 548 141($+$0) \\*[2ex]
5    & $-$1.996 450 584 458($-$2) & $-$2.797 226 479 764($-$1) & 1.065 244 131 032($+$2) & $-$1.068 841 233 198($+$2) & $-$2.398 615 568 695($-$1) \\
     & $-$1.996 450 584 455($-$2) & $-$2.797 226 479 762($-$1) & 1.065 244 130 018($+$2) & $-$1.068 841 232 184($+$2) & $-$2.398 615 568 694($-$1) \\*[2ex]
7    & $-$1.016 859 395 768($-$2) & $-$1.425 797 896 419($-$1) & 2.765 269 415 830($+$1) & $-$2.783 607 783 733($+$1) & $-$1.223 105 464 382($-$1) \\
     & $-$1.016 859 395 764($-$2) & $-$1.425 797 896 416($-$1) & 2.765 269 413 192($+$1) & $-$2.783 607 781 095($+$1) & $-$1.223 105 464 381($-$1) \\
\hline
\end{longtable}
\end{landscape}

\end{document}